%Paper: hep-th/9303019
%From: vtapia@halcon.dpi.udec.cl
%Date: Wed, 3 Mar 1993 15:03:03 +22311259 (CST)

\def\nt{\noindent}
\def\ce{\centerline}

\def\section#1#2{\vskip1truecm\nt{\bf#1\ #2}\vskip0.5truecm}\indent
\def\subsection#1#2{\vskip1truecm\nt{\bf#1\ #2}\vskip0.5truecm}\indent
\def\entry#1{\vskip1truecm\nt{\bf#1}\vskip0.5truecm}\indent

\magnification=1200
\normalbaselines
\nopagenumbers

\hsize=6.5 true in
\vsize=9.0 true in

\ce{\bf INTEGRABLE CONFORMAL FIELD THEORY IN FOUR DIMENSIONS}

\ce{\bf AND FOURTH-RANK GEOMETRY}

\bigskip

\ce{{\bf Victor Tapia}\footnote{*}{e-mail: VTAPIA@HALCON.DPI.UDEC.CL}}

\medskip

\ce{Departamento de F{\'\i}sica}

\ce{Facultad de Ciencias F{\'\i}sicas y Matem\'aticas}

\ce{Universidad de Concepci\'on}

\ce{Casilla 3-C}

\ce{Concepci\'on, Chile}

\bigskip

\nt{\bf Abstract.} We consider the conformal properties of geometries
described by higher-rank line elements. A crucial role is played by the
conformal Killing equation (CKE). We introduce the concept of null-flat
spaces in which the line element can be written as ${ds}^r=r!d\zeta_1\cdots
d\zeta_r$. We then show that, for null-flat spaces, the critical dimension,
for which the CKE has infinitely many solutions, is equal to the rank of the
metric. Therefore, in order to construct an integrable conformal field theory
in 4 dimensions we need to rely on fourth-rank geometry. We consider the
simple model ${\cal L}={1\over4}G^{\mu\nu\lambda\rho}
\partial_\mu\phi\partial_\nu\phi\partial_\lambda\phi\partial_\rho\phi$ and
show that it is an integrable conformal model in 4 dimensions. Furthermore,
the associated symmetry group is ${Vir}^4$.

\bigskip

\nt ICP. 04.50.+h Unified field theories and other theories of gravitation.

\vfill\eject

\section{1.}{Introduction}

It is expected that at very high energies physical processes become scale
invariant. In fact, in such regime all masses involved in any physical
process are small in comparison with the energies, and they can be put equal
to zero. There is therefore, no fundamental mass setting the scale of
energies, and therefore, all physical processes must be scale invariant.

The above speculation is confirmed by several experiments, as deep inelastic
scattering, which reveal that, at very high energies, physical phenomena
become scale invariant. Therefore, any field theory attempting to provide a
unified description of interactions must, at high energies, exhibit this
behaviour.

It can furthermore be shown, from a mathematical point of view, that scale
invariance implies conformal invariance.

Conformal field theories can be constructed in any dimension but only for
$d=2$ they exhibit a radically different behaviour. Chief among them is the
fact that there exists an infinite number of conserved quantities making the
theory an exactly solvable, or integrable, theory. The associated symmetry
group becomes infinite-dimensional and after convenient parametrisation of
its generators in terms of Fourier components is the familiar Virasoro
algebra. To be more precise the group is $Vir\oplus Vir$, with one Virasoro
algebra associated to each null space-time direction. Furthermore,
two-dimensional conformal integrable field theories hold several other
properties which are missing when formulated into higher-dimensional
space-times. Chief among them are those properties, such as
renormalisability, which make of its quantum field theoretical version a
mathematically consistent model. All the previous facts gave rise to the
success of string theories in the recent years; cf. ref. 1 for further
details.

Let us closer analise the situation. In a conformal field theory the symmetry
generators are the conformal Killing vectors. They are solutions of the
conformal Killing equation

$$\partial_\mu\xi_\nu\,+\,\partial_\nu\xi_\mu\,-\,{2\over d}\,g_{\mu\nu}\,
\partial_\lambda\xi^\lambda\,=\,0\,.\eqno{(1.1)}$$

\nt Only for two-dimensional spaces the solutions are infinitely many giving
rise to an infinite-dimensional symmetry group. A closer analysis of the
conformal Killing equation shows that this critical dimension is closely
related to the rank of the metric. In fact, since the metric is a second-rank
tensor, in the conformal Killing equation will appear two terms containing
derivatives of the Killing vectors. After contraction with the metric a 2 is
contributed which leads to the critical dimension $d=2$. Therefore, the
critical dimension for which the theory exhibits the critical behaviour is
equal to the rank of the metric.

However, 2 is quite different from 4, the accepted dimension of space-time.
Therefore, the ideal situation would be to have a field theory with the
previous properties holding in four dimensions: scale invariance, the
appearance of an infinite-dimensional symmetry group and the hope for a
mathematically consistent quantum field theory. Many attempts have been done
in order to reach this purpose. However, all existing proposals are not
exempt from criticism and no one can claim success. All of them have so far
met with considerable difficulties and in spite of the tremendous amount of
work done on the subject there is still not a generally accepted integrable
conformal field theory in four dimensions.

In order to obtain an integrable conformal field theory two ingredients are
necessary. The first one is the scale invariance and the second is the
existence of an infinite number of conserved quantities.

As mentioned above, scale invariance implies conformal invariance. However,
the concept of conformality strongly depends on the metric properties of the
base space. In order to clarify this point let us consider some general
properties of metric spaces.

Our approach is based on the use of line elements of the form

$${ds}^r\,=\,G_{\mu_1\cdots\mu_r}\,{dx}^{\mu_1}\,\cdots\,{dx}^{\mu_r}\,.
\eqno{(1.2)}$$

\nt Here the natural object is the rth-rank metric $G_{\mu_1\cdots\mu_r}$.
With it we can define generalised inner products, norms and angles.

The inner product of two generic vectors $A^\mu$ and $B^\mu$ is given by

$$N_{p,q}(A,\,B)\,=\,G_{\mu_1\cdots\mu_r}\,A^{\mu_1}\,\cdots\,A^{\mu_p}\,
B^{\mu_{p+1}}\,\cdots\,B^{mu_{p+q}}\,,\eqno{(1.3)}$$

\nt where, obviously, $p+q=r$. Next we can define the norm of the vector
$A^\mu$ by

$$A\,=\,{\mid N_{p,q}(A,\,A)\mid}^{1\slash r}\,=\,{\mid N_{r,0}(A,\,0)\mid}
^{1\slash r}\,,\eqno{(1.4)}$$

\nt and the same definition for $B^\mu$. Next we define the {\it generalised
angles}

$$\alpha_{p,q}(A,\,B)\,=\,A^{-p}\,B^{-q}\,N_{p,q}(A,\,B)\,.\eqno{(1.5)}$$

\nt In the second-rank case the above formulae coincide with the usual
definitions. These  formulae  seem  to  be  the  natural  generalisations  to
higher-rank geometries of the concepts of inner product, norm and angle, in
Riemannian geometry.

It can then be shown that the generalised angles are invariant under the
scale transformation

$$G_{\mu_1\cdots\mu_r}\,\rightarrow\,\Omega\,G_{\mu_1\cdots\mu_r}\,.
\eqno{(1.6)}$$

\nt Therefore, it is reasonable to call conformal this kind of
transformations. This is the concept of conformality we use in our approach.

The next step is to determine how conformal transformations of the metric
$G_{\mu_1\cdots\mu_r}$ can be obtained. For this let us consider the
transformations

$$x^\mu\,\rightarrow\,x^\mu\,+\,\xi^\mu\,,\eqno{(1.7)}$$

\nt with $\xi^\mu$ an infinitesimal function. At first order in $\xi^\mu$
the metric is changed by

$$\delta G_{\mu_1\cdots\mu_r}\,=\,G_{\mu_2\cdots\mu_r\mu}\,\partial_{\mu_1}
\xi^\mu\,+\,\cdots\,+\,G_{\mu_1\cdots\mu_{r-1}\mu}\,\partial_{\mu_r}\xi^\mu
\,+\,\xi^\mu\,\partial_\mu G_{\mu_1\cdots\mu_r}\,.\eqno{(1.8)}$$

\nt A conformal transformation will be induced on the metric if the previous
variation is proportional to the metric

$$\delta G_{\mu_1\cdots\mu_r}\,=\,\alpha\,G_{\mu_1\cdots\mu_r}\,.
\eqno{(1.9)}$$

\nt We obtain then a generalised conformal Killing equation

$$G_{\mu_2\cdots\mu_r\mu}\,\partial_{\mu_1}\xi^\mu+\cdots\,+\,G_{\mu_1\cdots
\mu_{r-1}\mu}\,\partial_{\mu_r}\xi^\mu\,+\,\xi^\mu\,\partial_\mu G_{\mu_1
\cdots\mu_r}$$

$$-\,{r\over d}\,G_{\mu_1\cdots\mu_r}\,G^{-1\slash r}\,\partial_\mu
(G^{1\slash r}\,\xi^\mu)\,=\,0\,,\eqno{(1.10)}$$

\nt (the value of $\alpha$ has been fixed by taking the trace of this
equation) where $G$ is the determinant of $G_{\mu_1\cdots\mu_r}$. Before
continuing the analysis of this equation let us turn our attention to the
curvature properties of differentiable manifolds.

Curvature properties are described by the curvature tensor

$${R^\lambda}_{\rho\mu\nu}\,=\,\partial_\mu{\Gamma^\lambda}_{\nu\rho}\,-\,
\partial_\nu{\Gamma^\lambda}_{\mu\rho}\,+\,{\Gamma^\lambda}_{\mu\sigma}\,{
\Gamma^\sigma}_{\nu\rho}\,-\,{\Gamma^\lambda}_{\nu\sigma}\,{\Gamma^\sigma}_{
\mu\rho}\,,\eqno{(1.11)}$$

\nt constructed in terms of a connection ${\Gamma^\lambda}_{\mu\nu}$.

The metric and the connection are, in general, independent objects. They can
be related through a metricity condition. The natural metricity condition is

$$\nabla_\mu G_{\mu_1\cdots\mu_r}\,=\,\partial_\mu G_{\mu_1\cdots\mu_r}\,-\,
{\Gamma^\lambda}_{\mu(\mu_1}\,G_{\mu_2\cdots\mu_r)}\,=\,0\,.\eqno{(1.12)}$$

\nt The number of unknowns for a symmetric connection ${\Gamma^\lambda}_
{\mu\nu}$ is ${1\over2}d^2(d+1)$, while the number of equations is

$${{d\,(d+r-1)!}\over{(d-1)!\,r!}}\,.\eqno{(1.13)}$$

\nt This number is always greater than ${1\over2}d^2(d+1)$. Therefore the
system is overdetermined and some differentio-algebraic conditions must be
satisfied by the metric. Since, in general, such restrictions will not be
satisfied by a generic metric, one must deal with the connection and the
metric as independent objects. Therefore, for physical applications, the
connection and the metric must be considered as independent fields.

The exception is $r=2$, Riemannian geometry, in which the number of unknowns
and the number of equations are the same. Since (1.12) is an algebraic linear
system, the solution is unique and is given by the familiar Christoffel
symbols of the second kind.

A metricity condition can be imposed consistently only if the number of
independent components of the metric is less than that naively implied by
(1.13).  The  maximum  acceptable  number  of   independent   components   is
${1\over2}d^2(d+1)$. This  can  be  achieved,  for  instance,  for  null-flat
spaces, for which the line element is given by

$${ds}^r\,=\,r!\,{d\zeta}^1\,\cdots\,{d\zeta}^r\,.\eqno{(1.14)}$$

\nt Spaces described by this line element have r null directions. The only
non-null component of the metric is

$$G_{1\cdots r}\,=\,1\,.\eqno{(1.15)}$$

\nt In this case eq. (1.12) has the  unique  solution  ${\Gamma^\lambda}_{\mu
\nu}=0$, therefore these spaces are flat. That is the  reason  to  call  them
{\it null-flat} spaces.

A simple counting of equations and unknowns shows that the situation for eq.
(1.10) is similar to that for eq. (1.12). Therefore a consistent solution
will exist only for certain kinds  of  metrics.  A  particular  case  is  for
null-flat metrics. In this case one can prove the following result:

\bigskip

{\bf Theorem.} {\it The critical dimension, for which the  conformal  Killing
equation has infinitely many solutions, is equal to the rank of the metric.}

\bigskip

\nt In this case one can furthermore prove that the symmetry group is
$Vir^r$.

Therefore, if we want to construct an integrable conformal field theory in
four dimensions we must rely on fourth-rank geometry. In this case the
conformal Killing equation must appear as the condition for the existence of
conserved quantities.

Let us now mimic the introductory remarks for a fourth-rank metric. Conformal
field theories can be constructed in any dimension but only for $d=4$ they
exhibit a radically different behaviour. Chief among them is the fact that
the symmetry group becomes infinite-dimensional. The group, after convenient
parametrisation of its generators in terms of Fourier components, is nothing
more than the familiar Virasoro algebra. To be more precise the group is
$Vir\oplus Vir\oplus Vir\oplus Vir$, with one Virasoro algebra for each null
space-time direction. The fact that the symmetry group is
infinite-dimensional implies that there is an infinite number of conserved
quantities making the theory an exactly solvable, or integrable, theory.
Furthermore, four-dimensional field theories hold several other properties
which are missing when formulated in other dimensions. Chief among them are
those properties which make of its quantum field theoretical version a
mathematically consistent model.

Let us closer analise the situation. In a conformal field theory the symmetry
generators are the conformal Killing vectors. They are solutions of the
conformal Killing equation

$$G_{\alpha\mu\nu\lambda}\,\partial_\rho\xi^\alpha\,+\,G_{\alpha\nu\lambda
\rho}\,\partial_\mu\xi^\alpha\,+\,G_{\alpha\lambda\rho\mu}\,\partial_\nu\xi^
\alpha\,+\,G_{\alpha\rho\mu\nu}\,\partial_\lambda\xi^\alpha\,-\,{4\over d}\,G
_{\mu\nu\lambda\rho}\,G^{-1\slash4}\,\partial_\alpha(G^{1\slash4}\,\xi^\alpha
)\,=\,0\,.\eqno{(1.16)}$$

\nt Only for four-dimensional spaces the solutions are infinitely many giving
rise to an infinite-dimensional symmetry group. A closer analysis of the
conformal Killing equation shows that this critical dimension is closely
related to the rank of the metric. In fact, since the metric is a fourth-rank
tensor, in the conformal Killing equation will appear four terms containing
derivatives of the Killing vectors. After contraction with the metric a 4 is
contributed which leads to the critical dimension $d=4$. Therefore, the
critical dimension for which the theory exhibits the critical behaviour is
equal to the rank of the metric.

Comparison of eq. (1.16) with eq. (1.1) illustrates the comments at the
introduction concerning the appearance of a number of terms equal to the rank
of the metric.

The simple Lagrangian

$${\cal L}\,=\,G^{\mu\nu\lambda\rho}\,\phi_\mu\,\phi_\nu\,\phi_\lambda\,\phi_
\rho\,G^{1\slash4}\,, \eqno{(1.17)}$$

\nt where $\phi_\mu=\partial_\mu\phi$, and $\phi$ is a scalar field, exhibits
all the properties we are looking for: it is conformally invariant and
integrable. Furthermore the Lagrangian (1.17) is renormalisable, by power
counting, in four dimensions.

The work is organised as follows: In Section 2 we start by considering the
metric properties of differentiable manifolds. In Section 3 we consider the
curvature properties of differentiable manifolds and introduce the concept of
null-flat spaces. Section 4 is dedicated to the conformal Killing equation in
null-flat spaces. In Section 5 we introduce the fundamentals of conformal
field theory. Section 6 reviews the results on conformal field theory for
second-rank, Riemannian, geometry. Section 7 presents the results on
conformal field theory for fourth-rank geometry. Section 8 is dedicated to
the conclusions.

To our regret, due to the nature of this approach, we must bore the reader by
exhibiting some standard and well known results in order to illustrate where
higher-rank geometry departs from the standard one.

\section{2.}{Metric Properties of Differentiable Manifolds}

The metric properties of a differentiable manifold are related to the way in
which one measures distances. Let us remember the fundamental definitions
concerning the metric properties of a manifold. Here we take recourse to the
classical argumentation by Riemann.$^2$

Let $M$ be a {\it d}-dimensional differentiable manifold, and let $x^\mu$,
$\mu=1,\cdots,d$, be local coordinates. The infinitesimal element of distance
$ds$ is a function of the coordinates $x$ and their differentials $dx$'s

$$ds\,=\,f(x,\,dx)\,,\eqno{(2.1)}$$

\nt which is homogeneous of the first-order in $dx$'s

$$f(x,\,\lambda\,dx)\,=\,\lambda\,f(x,\,dx)\,,\eqno{(2.2a)}$$

\nt $\lambda >0$, and is positive definite

$$f\,\geq\,0\,.\eqno{(2.2b1)}$$

\nt Condition (2.2b1) was written, so to say, in a time in which line
elements were thought to be positive definite. With the arrival of General
Relativity one got used to work with line elements with undefined signature.
Condition (2.2b1) was there to assure that the distance measured in one
direction is the same one measures in the opposite direction. Therefore,
condition (2.2b1) can be replaced by the weaker condition

$$f(x,\,-dx)\,=\,f(x,\,dx)\,.\eqno{(2.2b2)}$$

\nt However, the above conditions can be summarised into the single condition

$$f(x,\,\lambda\,dx)\,=\,\mid\lambda\mid\,f(x,\,dx)\,.\eqno{(2.2)}$$

Of course the possibilities are infinitely many. Let us restrict our
considerations to monomial functions. Then we will have

$$ds\,=\,{(G_{\mu_1\cdots\mu_r}(x)\,{dx}^{\mu_1}\,\cdots\,{dx}^{\mu_r})}
^{1\slash r}\,.\eqno{(2.3)}$$

\nt Condition (2.2a) is satisfied by construction. In order to satisfy
condition (2.2b2) $r$ must be an even number.

The simplest choice is $r=2$

$${ds}^2\,=\,g_{\mu\nu}\,dx^\mu\,dx^\nu\,,\eqno{(2.4)}$$

\nt which corresponds to Riemannian geometry. The coefficients $g_{\mu\nu}$
are the components of the covariant metric tensor. The determinant of the
metric is given by

$$g\,=\,{1\over{d!}}\,\epsilon^{\mu_1\cdots\mu_d}\,\epsilon^{\nu_1\cdots\nu_d
}\,g_{\mu_1\nu_1}\,\cdots\,g_{\mu_d\nu_d}\,.\eqno{(2.5)}$$

If $g\not=0$, the inverse metric is given by

$$g^{\mu\nu}\,=\,{1\over{(d-1)!}}\,{1\over g}\,\epsilon^{\mu\mu_1\cdots\mu_{d
-1}}\,\epsilon^{\nu\nu_1\cdots\nu_{d-1}}\,g_{\mu_1\nu_1}\,\cdots\,g_{\mu_{d-1
}\nu_{d-1}}\,.\eqno{(2.6)}$$

\nt and satisfies

$$g^{\mu\lambda}\,g_{\lambda\nu}\,=\,\delta^\mu_\nu\,.\eqno{(2.7)}$$

The next possibility is $r=4$. In this case the line element is given by

$${ds}^4\,=\,G_{\mu\nu\lambda\rho}\,{dx}^\mu\,{dx}^\nu\,{dx}^\lambda\,{dx}^
\rho\,.\eqno{(2.8)}$$

\nt The determinant of the metric $G_{\mu\nu\lambda\rho}$ is given by

$$G\,=\,{1\over{d!}}\,\epsilon^{\mu_1\,\cdots\mu_d}\,\cdots\,\epsilon^{\rho_1
\cdots\rho_d}\,G_{\mu_1\nu_1\lambda_1\rho_1}\,\cdots\,G_{\mu_d\nu_d\lambda_d
\rho_d}\,,\eqno{(2.9)}$$

\nt  where  the  $\epsilon$'s  can  be  chosen  as   the   usual   completely
antisymmetric Levi-Civita symbols. If $G\not= 0$, the inverse metric is given
by

$$G^{\mu\nu\lambda\rho}\,=\,{1\over{(d-1)!}}\,{1\over G}\,\epsilon^{\mu\mu_1
\cdots\mu_{d-1}}\,\cdots\,\epsilon^{\rho\rho_1\cdots\rho_{d-1}}\,G_{\mu_1
\nu_1\lambda_1\rho_1}\,\cdots\,G_{\mu_{d-1}\nu_{d-1}\lambda_{d-1}\rho_{d-1}}
\,,\eqno{(2.10)}$$

\nt and satisfies the relations

$$G^{\mu\alpha\beta\gamma}\,G_{\nu\alpha\beta\gamma}\,=\,\delta^\mu_\nu\,.
\eqno{(2.11)}$$

\nt That this is true can be verified by hand in the two-dimensional case and
with computer algebraic manipulation for 3 and 4 dimensions.$^3$

All the previous results can be generalised to an arbitrary even $r$. In the
generic case the line element is given by

$${ds}^r\,=\,G_{\mu_1\cdots\mu_r}\,{dx}^{\mu_1}\,\cdots\,{dx}^{\mu_r}\,.
\eqno{(2.12)}$$

\nt The determinant of the metric $G_{\mu_1\cdots\mu_r}$ is given by

$$G\,=\,{1\over d!}\,\epsilon^{\alpha_1\cdots\alpha_d}\,\cdots\,\epsilon^{
\rho_1\cdots\rho_d}\,G_{\alpha_1\cdots\rho_1}\,\cdots\,G_{\alpha_d\cdots\rho_
d}\,,\eqno{(2.13)}$$

\nt where again the $\epsilon$'s can be chosen as the usual completely
antisymmetric Levi-Civita symbols. If $G\not=0$, the inverse metric is given
by

$$G^{\alpha\cdots\rho}\,=\,{1\over{(d-1)!}}\,{1\over G}\,\epsilon^{\alpha
\alpha_1\cdots\alpha_{d-1}}\,\cdots\,\epsilon^{\rho\rho_1\cdots\rho_{d-1}}\,G
_{\alpha_1\cdots\rho_1}\,\cdots\,G_{\alpha_{d-1}\cdots\rho_{d-1}}\,,
\eqno{(2.14)}$$

\nt and satisfies the relations

$$G^{\mu\lambda_1\cdots\lambda_{r-1}}\,G_{\nu\lambda_1\cdots\lambda_{r-1}}\,=
\,\delta^\mu_\nu\,.\eqno{(2.15)}$$

It is clear that higher-rank geometries are observationally excluded at the
scale of distances of our daily life. However, a Riemannian behaviour can be
obtained for {\it separable spaces}. A space is said to be {\it separable} if
the metric decomposes as

$$G_{\mu_1\nu_1\cdots\mu_s\nu\_s}\,=\,g_{(\mu_1\nu_1}\,\cdots\,g_{\mu_s\nu_s)
}\,.\eqno{(2.16)}$$

\nt In this case the line element reduces to a quadratic form and therefore
all the results obtained for a generic $G_{\mu_1\cdots\mu_r}$ reduce to
those for Riemannian geometry.

\subsection{2.1.}{Inner Products and Angles}

Let us consider the inner products of two generic vectors $A^\mu$ and $B^\mu$

$$N_{p,q}(A,\,B)\,=\,G_{\mu_1\cdots\mu_r}\,A^{\mu_1}\,\cdots\,A^{\mu_p}\,B^{
\mu_{p+1}}\,\cdots\,B^{\mu_{p+q}}\,,\eqno{(2.17)}$$

\nt where, obviously, $p+q=r$. Next we can define the norm of the vector
$A^\mu$ by

$$A\,=\,{\mid N_{p,q}(A,\,A)\mid}^{1\slash r}\,=\,{\mid N_{r,0}(A,\,0)\mid}^{
1\slash r}\,,\eqno{(2.18)}$$

\nt and the same definition for $B^\mu$. Next we define the {\it generalised
angles}

$$\alpha_{p,q}(A,\,B)\,=\,A^{-p}\,B^{-q}\,N_{p,q}(A,\,B)\,.\eqno{(2.19)}$$

\nt In the second-rank case the above formulae coincide with the usual
definitions. These  formulae  seem  to  be  the  natural  generalisations  to
higher-rank geometries of the concepts of inner product, norm and angle, in
Riemannian geometry. Furthermore, it must be observed that for higher-rank
geometries we can consider inner products of more than two vectors. For our
purposes it is enough to restrict our considerations to two vectors.

Let us now observe that under the transformation

$$G_{\mu_1\cdots\mu_r}\,\rightarrow\,\Omega\,G_{\mu_1\cdots\mu_r}\,,
\eqno{(2.20)}$$

\nt the generalised angles remain unchanged, they are scale invariant. This
is a good reason to call the previous transformations {\it conformal}, since
they preserve the (generalised) angles.

\vfill\eject

\subsection{2.2.}{The Conformal Killing Equation}

Let us next analyse how one can obtain conformal symmetries of the metric.
Let us consider the transformation

$$x^\mu\,\rightarrow\,x^\mu\,+\,\xi^\mu(x)\,,\eqno{(2.21)}$$

\nt with $\xi$ an infinitesimal function. Under this transformation the
metric is changed, at first-order in $\xi$, by

$$\delta G_{\mu_1\cdots\mu_r}\,=\,G_{\mu_2\cdots\mu_r\mu}\,\partial_{\mu_1}
\xi^\mu\,+\,\cdots\,+\,G_{\mu_1\cdots\mu_{r-1}\mu}\,\partial_{\mu_r}\xi^\mu\,
+\,\xi^\mu\,\partial_\mu G_{\mu_1\cdots\mu_r}\,.\eqno{(2.22)}$$

\nt In order for this variation to induce a conformal transformation over the
metric, it must be

$$\delta G_{\mu_1\cdots\mu_r}\,=\,\alpha\,G_{\mu_1\cdots\mu_r}\,.
\eqno{(2.23)}$$

\nt One arrives then to the conformal Killing equation

$$G_{\mu_2\cdots\mu_r\mu}\,\partial_{\mu_1}\xi^\mu\,+\,\cdots\,+\,G_{\mu_1
\cdots\mu_{r-1}\mu}\,\partial_{\mu_r}\xi^\mu\,+\,\xi^\mu\,\partial_\mu G_{\mu
_1\cdots\mu_r}$$

$$-\,{r\over d}\,G_{\mu_1\cdots\mu_r}\,G^{-1\slash r}\,\partial_\mu(G^{1
\slash r}\,\xi^\mu)\,=\,0\,.\eqno{(2.24}$$

\nt (The value of $\alpha$ has been fixed by taking the trace of this
equation.) This equation is completely written in terms of the metric since
there is no Christoffel symbol associated to it. This equation is furthermore
overdetermined. In fact the number of derivatives $\partial_\nu\xi^\mu$ is
much lesser than the number of equations (2.24). Therefore, solutions will
exist only for certain classes of metrics. A solution can be obtained for
{\it null flat} spaces; cf. Section 3.1.

\section{3.}{Curvature Properties of Differentiable Manifolds}

Curvature properties are baseed on affine properties which in turn are
related to how one moves from one point to a close one. These properties are
mathematically described by the connection ${\Gamma^\lambda}_{\mu\nu}$. In
terms of the connection one can define the Riemann tensor

$${R^\lambda}_{\rho\mu\nu}\,=\,\partial_\mu{\Gamma^\lambda}_{\nu\rho}\,-\,
\partial_\nu{\Gamma^\lambda}_{\mu\rho}\,+\,{\Gamma^\lambda}_{\mu\sigma}\,{
\Gamma^\sigma}_{\nu\rho}\,-\,{\Gamma^\lambda}_{\nu\sigma}\,{\Gamma^\sigma}_{
\mu\rho}\,.\eqno{(3.1)}$$

Up  to  now  the  connection  ${\Gamma^\lambda}_{\mu\nu}$  and   the   metric
$G_{\mu\nu\lambda\rho}$  are  unrelated.  They  can  be  related  through   a
metricity condition. In Riemannian geometry this metricity condition reads

$$\nabla_\lambda g_{\mu\nu}\,=\,\partial_\lambda g_{\mu\nu}\,-\,{\Gamma^\rho}
_{\lambda\nu}\,g_{\mu\lambda}\,-\,{\Gamma^\rho}_{\lambda\mu}\,g_{\nu\lambda}
\,=\,0\,.\eqno{(3.2)}$$

\nt The number of unknowns  for  a  symmetric  $\Gamma$  and  the  number  of
equations (3.2) are the  same,  {\it  viz.}  ${1\over2}d^2(d+1)$.  Therefore,
since this is an algebraic linear system, the solution is unique and is given
by the familiar Christoffel symbols of the second kind

$${\Gamma^\lambda}_{\mu\nu}\,=\,\lbrace^\lambda_{\mu\nu}\rbrace(g)\,=\,{1
\over2}\,g^{\lambda\rho}\,(\partial_\mu g_{\nu\rho}\,+\,\partial_\nu g{\mu
\rho}\,-\,\partial_\rho g_{\mu\nu})\,.\eqno{(3.3)}$$

In the case of a higher-rank metric a condition analogous to (3.2) would read

$$\nabla_\mu G_{\mu_1\cdots\mu_r}\,=\,\partial_\mu G_{\mu_1\cdots\mu_r}\,-\,{
\Gamma^\lambda}_{\mu(\mu_1}\,G_{\mu_2\cdots\mu_r)\lambda}\,=\,0\,.
\eqno{(3.4)}$$

\nt However, in this case, the number of unknowns ${\Gamma^\lambda}_{\mu\nu}$
is, as before, ${1\over2}d^2(d+1)$, while the number of equations is

$$d\,{(d+r-1)!\over{r!(d-1)!}}\,>\,d\,{{d(d+1)}\over2}\,,\,for\,r>2\,.
\eqno{(3.5)}$$

\nt Therefore the system is overdetermined and some differentio-algebraic
conditions must be satisfied by the metric. Since, in general, such
restrictions will not be satisfied by a generic metric, one must deal with
${\Gamma^\lambda}_{\mu\nu}$ and $G_{\mu_1\cdots\mu_r}$ as independent
objects. A metricity condition can be imposed consistently only if the number
of independent components of the metric is lesser than that naively implied
by (3.5). The maximum acceptable number of independent components is
${1\over2}d(d+1)$. This can be achieved, for instance, if the metric is  that
corresponding to a null-flat one.

\subsection{3.1}{Null-Flat Spaces}

As we will see in the next section, there is a close connection between the
dimension of the manifold and the rank of the metric. For a second-rank
geometry in a two-dimensional manifold we have that the metric of a flat
space can always be brought to the forms

$${ds}^2\,=\,{dt}^2\,-\,{dx}^2\,,\eqno{(3.6a)}$$

$${ds}^2\,=\,{du}^2\,+\,{dv}^2\,,\eqno{(3.6b)}$$

\nt for Minkowskian and Euclidean signatures, respectively. However, both of
them can be brought to the simple form

$${ds}^2\,=\,2\,dz^+\,dz^-\,.\eqno{(3.7)}$$

\nt where

$$z^\pm\,=\,{1\over{\sqrt2}}\,(t\,\pm\,x)\,,\eqno{(3.8a)}$$

$$z^\pm\,=\,{1\over{\sqrt2}}\,(u\,\pm\,i\,v)\,,\eqno{(3.8b)}$$

\nt for the Minkowskian and Euclidean cases, respectively. Therefore the
canonical form (3.7) is independent of the signature of the underlying space.
Since for higher-rank geometries there is no concept of flatness, eq. (3.7)
seems to be a good definition to be generalised.

The concept of {\it null-flat} space is defined only for spaces in which the
dimension and the rank of the metric coincide. Then, the line element is
given by

$${ds}^r\,=\,r!\,{d\zeta}^1\,\cdots\,{d\zeta}^r\,.\eqno{(3.9)}$$

\nt (In higher-rank geometry not all flat spaces are null.) It is clear that
each coordinate $\zeta_\mu$ is associated to a null direction of the
manifold. This is the reason to call these spaces {\it null}. The only
non-null component of the metric is

$$G_{1\cdots r}\,=\,1\,.\eqno{(3.10)}$$

\nt One can now easily verify that in this case the metricity condition for
higher-rank metrics, eq. (3.4), has the only solution
${\Gamma^\lambda}_{\mu\nu}=0$. Therefore these spaces are flat. This is the
reason to call these spaces {\it flat}.

\section{4.}{The Conformal Killing Equation in Null-Flat Spaces}

Now we come to what we consider to be our most important result. Let us
consider the conformal Killing equation in a null-flat space

$$G_{\mu_2\cdots\mu_r\mu}\,\partial_{\mu_1}\xi^\mu\,+\,\cdots\,+\,G_{\mu_1
\cdots\mu_{r-1}\mu}\,\partial_{\mu_r}\xi^\mu\,-\,G_{\mu_1\cdots\mu_r}\,
\partial_\mu\xi^\mu\,=\,0\,.\eqno{(4.1)}$$

\nt (The ${r\over d}$ factor has  disappeared  since  in  a  null-flat  space
$r=d$.) Then we can establish the following:

\bigskip

{\bf Theorem.} {\it The critical dimension, for which the conformal Killing
equation has infinitely many solutions, is equal to the rank of the metric.}

\bigskip

The proof is quite simple. Let us observe that at most two indices can be
equal. The number of equations in which two indices are equal is $d(d-1)$.
They are equivalent to

$$\partial_\nu\xi^\mu\,=\,0\,,\,\nu\,\not=\,\mu\,.\eqno{(4.2)}$$

\nt The solution is (no summation)

$$\xi^\mu\,=\,f^\mu(\zeta^\mu)\,.\eqno{(4.3)}$$

\nt The equation in which all indices are different is identically zero.
Therefore, the components of the conformal Killing vectors are arbitrary
functions of the single coordinate along the associated direction and
therefore the solutions are infinitely many.

Let us now define the operators (no summation)

$$U^\mu(f)\,=\,\xi^\mu(\zeta^\mu)\,\partial_\mu\,=\,f^\mu(\zeta^\mu)\,
\partial_\mu\,.\eqno{(4.4)}$$

\nt One can then easily verify that

$$\lbrace U^\mu(f_1),\,U^\nu(f_2)\rbrace\,=\,U^\mu(f_1)\,U^\nu(f_2)\,-\,U^\nu
(f_2)\,U^\mu(f_1)\,=\,\delta^{\mu\nu}\,U(f_1\,{f_2}'\,-\,f_2\,{f_1}')\,.
\eqno{(4.5)}$$

\nt Therefore the symmetry group is the direct product of $r$ times the group

$$\lbrace U(f_1),\,U(f_2)\rbrace\,=\,U(f_1)\,U(f_2)\,-\,U(f_2)\,U(f_1)\,.
\eqno{(4.6)}$$

\nt which, after Fourier parametrisation, we recognise as the Virasoro group.
The symmetry group is therefore ${Vir}^r$.

\section{5.}{Integrable Conformal Field Theories}

Now we state the fundamentals for the construction of an integrable conformal
field theory.

Let us start by making some elementary considerations about field theory. We
will restrict our considerations to generic fields $\phi^A$,  $A=1,\cdots,n$,
described by a Lagrangian

$${\cal L}\,=\,{\cal L}(\phi^A,\,{\phi^A}_\mu)\,,\eqno{(5.1)}$$

\nt where ${\phi^A}_\mu=\partial_\mu\phi^A$. The field equations are

$${{\delta{\cal L}}\over{\delta\phi^A}}\,=\,{{\partial{\cal L}}\over{\partial
\phi^A}}\,-\,d_\mu{\pi_A}^\mu\,=\,0\,,\eqno{(5.2)}$$

\nt where we have introduced the generalised canonical momentum

$${\pi_A}^\mu\,=\,{{\partial{\cal L}}\over{\partial{\phi^A}_\mu}}\,.
\eqno{(5.3)}$$

\nt The energy-momentum tensor is given by

$${{\cal H}^\mu}_\nu\,=\,{\phi^A}_\nu\,{\pi_A}^\mu\,-\,\delta^\mu_\nu\,{\cal
L}\,,\eqno{(5.4)}$$

\nt and satisfies the continuity equation

$$d_\mu{{\cal H}^\mu}_\nu\,=\,-\,{\phi^A}_\nu\,{{\delta{\cal L}}\over{\delta
\phi^A}}\,=\,0\,.\eqno{(5.5)}$$

\nt The first comment relevant to our work is in order here. The definition
(5.4), of the energy-momentum tensor, guarantees, through (5.5), its
conservation on-shell. This definition is independent of the existence of a
metric or other background field. This is what we need in the next stages
where we are going to independise from the usual second-rank metric.

Let us now make some considerations about conformal field theory. The main
properties that a conformal theory must have are:
\item{C1.} Translational invariance, which implies that the energy-momentum
tensor ${{\cal H}^\mu}_\nu$ is conserved, i.e., eq. (5.5).
\item{C2.} Invariance under scale transformations which implies the existence
of the dilaton current. This current can be constructed to be$^4$

$$D^\mu\,=\,{{\cal H}^\mu}_\nu\,x^\nu\,.\eqno{(5.6)}$$

\nt  The  conservation  of  $D^\mu$  implies  that  ${{\cal  H}^\mu}_\nu$  is
traceless

$$d_\mu D^\mu\,=\,{{\cal H}^\mu}_\mu\,=\,0\,,\eqno{(5.7)}$$

\nt where the conservation of ${{\cal H}^\mu}_\nu$, eq. (5.5), has been used.

Now we look for the possibility of constructing further conserved quantities.
We concentrate on quantities of the form

$$J^\mu\,=\,{{\cal H}^\mu}_\nu\,\xi^\nu\,.\eqno{(5.8)}$$

\nt Then it must be

$$d_\mu J^\mu\,=\,{{\cal H}^\mu}_\nu\,d_\mu\xi^\nu\,=\,0\,.\eqno{(5.9)}$$

\nt In order to obtain more information from this equation we need to
introduce a further geometrical object allowing us to raise and low indices.

In the next sections we apply the above condition to second- and fourth-rank
geometries.

\section{6.}{Scale Invariant Field Theory in Second-Rank Geometry}

Now we consider the properties of scale invariant field theories in
second-rank geometry.

In order to obtain the consequences of scale invariance in second-rank
geometry we consider a constant flat metric $g_{\mu\nu}$. Then we define

$${\cal H}^{\mu\nu}\,=\,{{\cal H}^\mu}_\lambda\,g^{\lambda\nu}\,,
\eqno{(6.1)}$$

$$\xi_\mu\,=\,g_{\mu\nu}\,\xi^\nu\,.\eqno{(6.2)}$$

\nt If (6.1) happens to be symmetric then eq. (5.9) can be written as

$$d_\mu J^\mu\,=\,{1\over2}\,{\cal H}^{\mu\nu}\,(\partial_\mu\xi_\nu\,+\,
\partial_\nu\xi_\mu)\,=\,0\,.\eqno{(6.3)}$$

\nt Furthermore, if the energy-momentum tensor is traceless, as required by
scale invariance, the most general solution to (6.3) is

$$\partial_\mu\xi_\nu\,+\,\partial_\nu\xi_\mu\,-\,{2\over d}\,g_{\mu\nu}\,
\partial_\lambda\xi^\lambda\,=\,0\,.\eqno{(6.4)}$$

\nt i.e., the $\xi$'s are conformal Killing vectors for the metric
$g_{\mu\nu}$.

As shown in section 4, the critical dimension for this equation is $d=2$,
i.e., the solutions to (6.4) are infinitely many. For the metric (3.7) the
solutions to eq. (6.4) are

$$\xi^+\,=\,f(z^+)\,,\eqno{(6.5a)}$$

$$\xi^-\,=\,g(z^-)\,,\eqno{(6.5b)}$$

\nt where $f$ and $g$ are arbitrary functions. Now we define

$$U_+(f)\,=\,\xi^+\,\partial_+\,=\,f(z^+)\,\partial_+\,,\eqno{(6.6a)}$$

$$U_-(g)\,=\,\xi^-\,\partial_-\,=\,g(z^-)\,\partial_-\,.\eqno{(6.6b)}$$

\nt These quantities satisfy the commutation relations

$$\lbrace U_+(f_1),\,U_+(f_2)\rbrace\,=\,U_+(f_1\,{f_2}'\,-\,f_2\,{f_1}')\,,
\eqno{(6.7a)}$$

$$\lbrace U_+(f),\,U_-(g)\rbrace\,=\,0\,,\eqno{(6.7b)}$$

$$\lbrace U_-(g_1),\,U_-(g_2)\rbrace\,=\,U_-(g_1\,{g_2}'\,-\,g_2\,{g_1}')\,.
\eqno{(6.7c)}$$

\nt Relations (6.7) are essentially the algebra of two-dimensional
diffeomorphisms. After conveniently parametrise them in terms of Fourier
components one gets the familiar Virasoro algebra. To be more precise there
is one Virasoro algebra for each null direction.

The next step is to find a conformal field theory for which the conserved
quantities be determined by eq. (6.4). The simplest example is

$${\cal L}\,=\,{1\over2}\,g^{\mu\nu}\,\phi_\mu\,\phi_\nu\,g^{1\slash2}\,,
\eqno{(6.8)}$$

\nt where $\phi_\mu=\partial_\mu\phi$. This simple Lagrangian has many
important properties. It is conformally invariant, it possesses infinitely
many conserved quantities and is renormalisable, by power counting, for
$d=2$. The energy-momentum tensor is

$${{\cal H}^\mu}_\nu\,=\,\phi_\nu\,g^{\mu\lambda}\,\phi_\lambda\,g^{1\slash2}
\,-\,\delta^\mu_\nu\,{\cal L}\,.\eqno{(6.9)}$$

\nt The contravariant form is

$${\cal H}^{\mu\nu}\,=\,(g^{\nu\rho}\,\phi_\rho)\,(g^{\mu\lambda}\,\phi_
\lambda)\,g^{1\slash2}\,-\,g^{\mu\nu}\,{\cal L}\,.\eqno{(6.10)}$$

\nt In this case the contravariant $\partial\phi$ coincides with the momentum
such that this expression becomes symmetric. It is furthermore traceless.
Therefore we will have an infinite set of conserved quantities, for $d=2$.

The above properties of the CKE in $d=2$ is the origin of the great success
of string theories. In fact, string theories have all the good properties one
would like for a consistent quantum field theory of the fundamental
interactions. All these reasons gave hope for strings to be the theory of
everything. However, string theories are interesting only when formulated in
2 dimensions, which is quite different from 4, the accepted dimension of
space-time. Therefore, the ideal situation would be to have a field theory
formulated in 4 dimensions and exhibiting the good properties of string
theories. This is the problem to which we turn our attention now.

\section{7.}{Integrable Conformal Field Theory in Four Dimensions}

As mentioned previously, in order to obtain an integrable scale invariant
model, two ingredients are necessary: scale invariance and an infinite number
of conserved quantities.

In order to obtain an infinite number of conserved quantities we need that
they be generated by an equation admitting infinitely many solutions.

In 4 dimensions an infinite number of solutions can be obtained with the
conformal Killing equation for fourth-rank geometry in null-flat spaces, as
shown in section 4. Now we need to establish the equivalence between the
condition (5.9) and the fourth-rank CKE

$$G_{\alpha\mu\nu\lambda}\,\partial_\rho\xi^\alpha\,+\,G_{\alpha\nu\lambda
\rho}\,\partial_\mu\xi^\alpha\,+\,G_{\alpha\lambda\rho\mu}\,\partial_\nu\xi^
\alpha\,+\,G_{\alpha\rho\mu\nu}\,\partial_\lambda\xi^\alpha\,-\,{4\over d}\,G
_{\mu\nu\lambda\rho}\,G^{-1\slash4}\,\partial_\alpha(G^{1\slash4}\,\xi^\alpha
)\,=\,0\,.\eqno{(7.1)}$$

In the fourth-rank case, however, the operation of raising and lowering
indices is not well defined, therefore the operations involved in (6.1)-(6.2)
do not exist. In fact this procedure works properly for second-rank metrics
due to the fact that only for them the operations of raising and lowering
indices are well defined (the tangent and cotangent bundles are
diffeomorphic). This is not a real problem since the only thing we must
require is that (7.1) gives rise to the conformal Killing equation. This can
be done for a simple Lagrangian which is the natural generalisation of (6.8)
to fourth-rank geometry.

Let us consider the Lagrangian

$${\cal L}\,=\,{1\over4}\,G^{\alpha\beta\gamma\delta}\,\phi_\alpha\,\phi_
\beta\,\phi_\gamma\,\phi_\delta\,G^{1\slash4}\,.\eqno{(7.2)}$$

\nt This simple Lagrangian exhibits the properties we are interested in. It
is conformally invariant, it possesses infinitely many conserved quantities
and it is renormalisable, by power counting, for $d=4$.

The generalised momenta are given by

$$\pi^\mu\,=\,G^{\mu\beta\gamma\delta}\,\phi_\beta\,\phi_\gamma\,\phi_\delta
\,G^{1\slash4}\,.\eqno{(7.3)}$$

\nt The energy-momentum tensor is defined as in (5.4). Condition (5.9) reads

$$\phi_\nu\,G^{\mu\beta\gamma\delta}\,\phi_\beta\,\phi_\gamma\,\phi_\delta\,
\partial_\mu\xi^\nu\,-\,{1\slash4}\,G^{\alpha\beta\gamma\delta}\,\phi_\alpha
\,\phi_\beta\,\phi_\gamma\,\phi_\delta\,\partial_\mu\xi^\mu\,=\,0\,.
\eqno{(7.4)}$$

\nt Since the $\xi$'s do not depend on $\partial\phi$'s, what must be zero is
the completely symmetric coefficient with respect to $\partial\phi$'s. The
result is the conformal Killing equation (7.1). Therefore, we will have an
infinite-dimensional symmetry group for $d=4$.

Therefore, we have suceeded in implementing conformal invariance for $d=4$.
We have seen furthermore that the rank of the metric is essential to
implement conformal invariance in higher dimensions. It must furthermore be
observed that the appearance of the conformal behaviour for some critical
dimension is a geometrical property of the base space and therefore it is
model independent. Therefore any attempt at the implementation of conformal
invariance in four dimensions by relying only on the second-rank metric is
condemned to fail.

The next step is of course to construct a more realistic model on lines, for
example, similar to the Polyakov string.

\subsection{7.1.}{Comments}

That the symmetry group for an integrable conformal  model  in  4  dimensions
should be ${Vir}^4$ was advanced by Fradkin and Linetsky.$^5$ While  Vir  and
${Vir}^2$ are clearly related to Riemannian geometry, a  similar  geometrical
concept was lacking for ${Vir}^4$. This missing geometrical link is  provided
by fourth-rank geometry. According to  our  previous  results  the  conformal
Killing equation for a fourth-rank metric exhibits the desired behaviour. Our
problem is therefore reduced  to  construct  a  field  theory  in  which  the
conformal Killing equation plays this central role.

\section{8.}{Conclusions}

We have seen that integrable conformal field theories can be constructed in 4
dimensions if one relies on fourth-rank geometry. Furthermore, all desirable
properties of a quantum field theory are present when using fourth-rank
geometry, {\it viz.}, renormalisability (by power counting), integrability,
etc. There seems to be a close connection between the dimension and the rank
of the geometry. When they coincide, as happens for null-flat spaces, all
good properties show up. The why this is so is a question still waiting for
an answer.

Our future plan of work is to develop further models, even realistic, with
the previous properties which perhaps will provide clues to answer the
question asked in the above paragraph.

\entry{Acknowledgements}

This work has been possible thank to the hospitality of the Laboratory of
Theoretical Physics, Joint Institute for Nuclear Research, Dubna, the
Istituto di Fisica Matematica "J.-Louis Lagrange", Universit\`a di Torino,
and the International Centre for Theoretical Physics, Trieste. The work has
been much enriched, at different stages, by talks with M. Ferraris, M.
Francaviglia, P. Aichelburg, A.L. Marrakchi and P. Minning.

\vfill\eject

\entry{References}

\item{1.} M. B. Green, J. H. Schwarz and E. Witten, {\it Superstrings}
(Cambridge University Press, Cambridge, 1987)
\item{2.} B. Riemann, {\it \"Uber die Hypothesen welche, der Geometrie zu
Grunde liegen}, {\it Abh. K\"onigl. Gesellsch. Wiss. G\"ottingen} {\bf 13},
1 (1868).
\item{3.} M. Ferraris, private communication (1991).
\item{4.} J. Polchinski, {\it Nucl. Phys. B} {\bf 303}, 226 (1988).
\item{5.} E. S. Fradkin and V. Ya. Linetsky, {\it Phys. Lett. B} {\bf 253},
97 (1991).

\bye